\documentclass[10pt,conference]{IEEEtran}
\IEEEoverridecommandlockouts

\usepackage{cite}
\usepackage{amsmath,amssymb,amsfonts,amsthm}
\usepackage{graphicx}
\usepackage{textcomp}
\usepackage{xcolor}
\usepackage{mathtools}
\usepackage{stfloats}
\usepackage{multirow}
\usepackage{hyperref}
\usepackage{bm}
\usepackage{comment}
\usepackage{soul}
\usepackage{float}
\usepackage{subfigure}

\usepackage{tikz}
\usetikzlibrary{decorations.pathreplacing,matrix}
\usepackage[absolute,overlay]{textpos}

\usepackage{algorithm}
\usepackage[noend]{algorithmic}

\usepackage{mathabx}

\newcommand\norm[1]{\left\lVert#1\right\rVert}

\theoremstyle{definition}
\newtheorem{Theorem}{Theorem}
\newtheorem{Definition}{Definition}
\newtheorem{Problem}{Problem}

\newcommand{\tabincell}[2]{\begin{tabular}{@{}#1@{}}#2\end{tabular}}

\def\BibTeX{{\rm B\kern-.05em{\sc i\kern-.025em b}\kern-.08em
    T\kern-.1667em\lower.7ex\hbox{E}\kern-.125emX}}
\begin{document}

\title{Partial Equivalence Checking of Quantum Circuits}

\author{\IEEEauthorblockN{Tian-Fu Chen$^{1,2,3}$, Jie-Hong R. Jiang$^{1,2,3,4}$, and Min-Hsiu Hsieh$^{5}$}
\IEEEauthorblockA{$^1$\textit{Department of Electrical Engineering, National Taiwan University}, Taipei, Taiwan}
\IEEEauthorblockA{$^2$\textit{Graduate School of Advanced Technology, National Taiwan University}, Taipei, Taiwan}
\IEEEauthorblockA{$^3$\textit{Center for Quantum Science and Engineering,
National Taiwan University}, Taipei, Taiwan}
\IEEEauthorblockA{$^4$\textit{Graduate Institute of Electronics Engineering,
National Taiwan University}, Taipei, Taiwan}
\IEEEauthorblockA{$^5$\textit{Quantum Computing Research Center, Hon Hai Research Institute}, Taipei, Taiwan}
\IEEEauthorblockA{ghdftff542@gmail.com, jhjiang@ntu.edu.tw, min-hsiu.hsieh@foxconn.com}
}

\maketitle

\begin{abstract}
Equivalence checking of quantum circuits is an essential element in quantum program compilation, in which a quantum program can be synthesized into different quantum circuits that may vary in the number of qubits, initialization requirements, and output states.
Verifying the equivalences among the implementation variants requires proper generality. 
Although different notions of quantum circuit equivalence have been defined, prior methods cannot check observational equivalence between two quantum circuits whose qubits are partially initialized, which is referred to as \emph{partial equivalence}.
In this work, we prove a necessary and sufficient condition for two circuits to be partially equivalent. Based on the condition, we devise algorithms for checking quantum circuits whose partial equivalence cannot be verified by prior approaches.
Experiment results confirm the generality and demonstrate the efficiency and effectiveness of our method.
Our result may unleash the optimization power of quantum program compilation to take more aggressive steps.



\end{abstract}


\section{Introduction}
Equivalence checking plays an important role in the design flow of modern integrated circuits. 
It ensures the transformations done in the corresponding synthesis steps do not introduce errors.
Many scalable techniques, such as random simulation, satisfiability solving, decision diagrams, and structural similarity detection, have been developed and widely applied in classical circuit verification \cite{ceq3, ceq4, ceq1, ceq2, int, ic3, pdr}.
Due to the rapid progress of hardware and software developments in quantum computation, quantum circuit verification, especially equivalence checking, arises as an important issue in quantum program compilation.
Although practical quantum circuits remain relatively small until now, it remains challenging to their 
equivalence checking due to extraordinary quantum properties. 
Moreover, as quantum computing technology evolves, the size of quantum circuits will continue growing. 
Efficient equivalence checking tools for quantum circuits are essential.

\begin{figure}[t]
\centerline{\includegraphics[width=230pt]{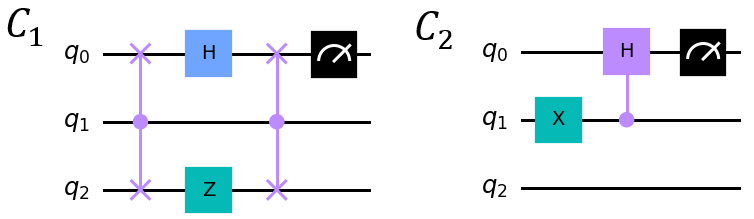}} 
\caption{A motivating example of partial equivalence where circuits $C_1$ and $C_2$ measured on $q_0$ have the same probability distribution over the outcomes for any given initial input state.} \label{partialExample}
\end{figure}

Many approaches to equivalence checking of quantum circuits have been proposed \cite{SliQEC,ref2,ref3,ref4,ref5,ref6,ref7,ref8,ref9,ref10,ref11,ref12}. 
Among them, decision diagrams are a widely adopted data structure for quantum state and operator representation and manipulation \cite{SliQEC,ref2,ref3,ref6,ref12}.
Also, techniques, such as miter construction \cite{SliQEC,ref3,ref5,ref6,ref11} or tensor network computation \cite{ref4}, have been applied. 
Some of them focus on the equivalence of reversible circuits, which realize permutation functions, whereas others address the equivalence of more general quantum circuits. 
For the latter, most of them require \emph{total equivalence}, namely, the identity of the output states (modulo a global phase difference) between the two circuits under verification.
However, since measurement is the only way to extract information from a quantum system, total quantum state equivalence may be unnecessary and only observational equivalence matters. That is, two circuits exhibit the same probability for every possible measurement outcome under a fixed measurement basis.
Furthermore, not all qubits of a circuit need to be measured and there can be a set of initial states to be verified not just a particular single initial state. 
To accommodate such different design constraints, we relax the notion of quantum circuit equivalence and define a generalization, called \emph{partial equivalence}. 
That is, two circuits are \emph{partially equivalent} if, given any valid initial input state, they exhibit the same probability for each measurement outcome.
Therefore, two partially equivalent circuits do not exhibit statistical differences by observed measurements.
Essentially, the generality of partial equivalence allows more flexibility for quantum circuit synthesis. 
Fig.~\ref{partialExample} shows an example where $C_1$ and $C_2$ are partially equivalent, but not totally equivalent. 

Apart from its application in quantum circuit synthesis, partial equivalence checking has its natural application in quantum program compilation. 
As the quantum computing technology evolves, many quantum programming languages have been developed, e.g., \cite{ref13,ref14,ref15,ref16,ref17,ref18,ref19,ref20,ref21,ref22,ref23,ref24}. 
Although nowadays most quantum programming languages are abstracted at the gate level,
high-level languages will get their increasing presence in the future. 
Their compilation may require making trade-offs between ancilla bits and quantum gates.
To verify the correctness of different compilation choices may require the defined partial equivalence.

Equivalence based on measurement results is not new, nor is the relaxation requiring only some of the input qubits to be data qubits and some of the output qubits to be measured qubits. 
In fact, partial equivalence can be seen as the case that all principal output qubits in Definition~3 of \cite{ref25} are to be measured.
However, there are no corresponding checking methods so far, and none of the prior work \cite{ref11,ref12,ref25,ref26,ref27} can fully resolve the partial equivalence checking problem.
In particular, equivalence up to a relative phase discussed in \cite{ref11} can only cover the case that all output qubits are measured; the method considering measurement-based equivalence in \cite{ref12} is used for reversible circuits only; the initial state is given and fixed in \cite{ref25} and \cite{ref26}; sequential circuit verification in \cite{ref27} is limited to the assumption that the number of measured output qubits cannot be smaller than the number of data input qubits.
In this work, we are not constrained by these limitations. 
Our methods can be applied to general cases of checking partial equivalence relations. 
Moreover, they are compatible with the state-of-the-art \textit{SliQEC} \cite{SliQEC} system, which uses Boolean functions represented by decision diagrams to store and manipulate matrices.
 

The main results of this work include:
\begin{enumerate}
\item We characterize \emph{partial equivalence} (in Section~\ref{sec:PECProblem}), which subsumes prior notions of quantum circuit equivalence. 
\item We prove a necessary and sufficient condition for two circuits to be partially equivalent (in Section~\ref{sec:theorem}).
\item We develop algorithms for partial equivalence checking under general and special settings (in Section~\ref{sec:alg}).
\item We conduct experiments to demonstrate the feasibility and effectiveness of our algorithms and the practical applicability of partial equivalence (in Section~\ref{sec:experiment}).
\end{enumerate}

\section{Preliminaries}
We briefly provide some background and define notations.

\subsection{Boolean Function and Decision Diagram}
For Boolean connectives, we denote conjunction by ``$\wedge$'' (sometimes omitted in a Boolean expression for brevity), 
disjunction by ``$\vee$,'' 
exclusive-or by ``$\oplus$,''
and negation by an overline.
Given a Boolean function $f(x_1,x_2,...,x_n)$, the \emph{positive cofactor} of $f$ on $x_i$, denoted $f|_{x_i}$, is
\begin{equation}\label{positiveCofactor} 
f|_{x_i} = f(x_1,x_2,...,x_{i-1},1,x_{i+1},...,x_n).
\end{equation}
Similarly, the \emph{negative cofactor} of $f$ on $x_i$, denoted $f|_{\overline{x_i}}$, is
\begin{equation}\label{negativeCofactor} 
f|_{\overline{x_i}} = f(x_1,x_2,...,x_{i-1},0,x_{i+1},...,x_n).
\end{equation}

\begin{figure}[H]
\centerline{\includegraphics[width=120pt]{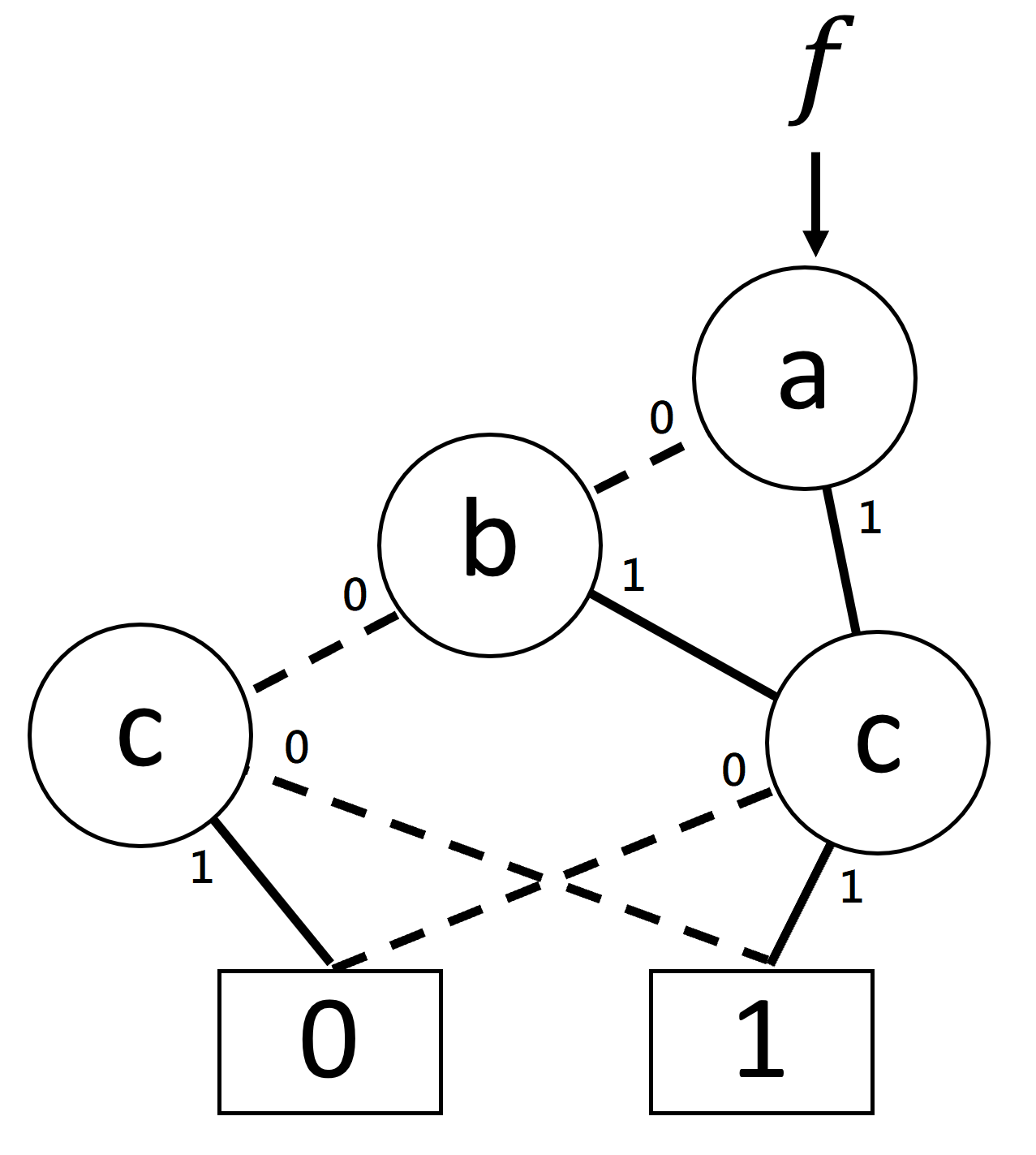}}
\caption{A BDD of function $f=ac \lor \overline{a}bc \lor \overline{a}\overline{b}\overline{c}$.
}\label{BDD}
\end{figure}

The \emph{reduced ordered binary decision diagram} (ROBDD), referred to as BDD in the sequel, is a canonical form for Boolean function representation \cite{ROBDD}.
There are efficient BDD packages for Boolean function manipulation.
A BDD is a directed acyclic graph consisting of non-terminal nodes and terminal-0 and terminal-1 nodes.
Each non-terminal node $v$ is associated with a decision variable and has two children $v_0$ and $v_1$ pointed to by the 0-branch and 1-branch of $v$, respectively.
Each node in a BDD corresponds to a Boolean function.
Let $f_v, f_{v_0}, f_{v_1}$ be the functions of node $v$ with decision variable $x$ and its two children $v_0, v_1$, respectively.
Then $f_{v_0} = f|_{\overline{x}}$ and $f_{v_1} = f|_{x}$, and the Shannon expansion $f_v = x f_{v_1} \vee \overline{x} f_{v_0}$ holds.
Fig.~\ref{BDD} shows a BDD of function $f= ac \vee \overline{a}bc \vee \overline{a} \overline{b} \overline{c}$.


\subsection{Quantum States and Quantum Circuits}

For a quantum system, at the input end, a qubit is called a \emph{data qubit} if it is associated with the input data and a \emph{non-data qubit} otherwise. 
At the output end, a qubit is called a \emph{measured qubit} if it is to be measured and a \emph{non-measured qubit} otherwise. 
In an $n$-qubit quantum system, a quantum state can be represented by a $2^n \times 1$ vector $\phi = [a_0, a_1, ..., a_{2^n-1}]^{\rm T}$, where each $a_i$ is a complex number and the norm of $\phi$ satisfies $\norm{\phi}=1$. 
We use $|0\rangle$ to represent the vector $[1,0,...,0]^{\rm T}$, $|1\rangle$ to represent the vector $[0,1,0,...,0]^{\rm T}$, and so on. 
For state $|i\rangle$, we let the binary number of integer $i$ be expressed by qubits $q_0, \ldots, q_{n-1}$ with $q_0$ being the most significant bit.
In the sequel, we assume that non-data qubits are fixed to the initial state $|0\rangle$ and refer to them as \emph{ancilla qubits}. 
This assumption is general in that any other initial state can be transformed from $|0\rangle$.

If we measure all the qubits of $\phi$ under the computational basis, there will be a probability ${|a_i|}^2$ for the system to get into state $|i\rangle$.
Also, if we measure only the first (more significant) $m$ qubits, for $m \leq n$, of $\phi$, the probability of obtaining a state $|j\rangle$, for $j \in \{0,1,...,2^m-1\}$, of the measured qubits equals
\begin{equation}
\sum_{k=g\cdot j}^{g\cdot (j+1)-1} {|a_k|}^2,
\end{equation}
\noindent 
where $g=2^{n-m}$.

The evolution of the state of a closed quantum system can be described by quantum gates. 
For an $n$-qubit quantum system, a quantum gate can be defined by a $2^n \times 2^n$ unitary matrix $U$ of complex numbers. 
A unitary matrix $U$ is a matrix satisfying $U^{\dagger}U=I$, where $U^{\dagger}$ is the conjugate transpose of $U$ and $I$ is the identity matrix. 
Thus, $U^{\dagger}$ is also the inverse of $U$. 
A quantum circuit is made up of a series of quantum gates $G_1,G_2,...G_d$.
Let their corresponding unitary matrices be $U_1, U_2, ..., U_d$ and let $\phi_0$ be the initial state. 
Then the final output state $\phi_{out}$ of the circuit is derived by the product
\begin{equation}
\phi_{out} = U_d\cdot U_{d-1}\cdot \ldots \cdot U_1\cdot \phi_0.
\label{eq1}\end{equation}
Therefore, we can view the whole circuit as a quantum gate with the unitary matrix
\begin{equation}
U = U_d\cdot U_{d-1}\cdot ...\cdot U_1.
\label{eq2}\end{equation}
Particularly, when an $n$-qubit quantum circuit with unitary matrix $U$ yields a permutation of the computational basis $\{|0\rangle, ..., |2^n-1\rangle\}$, it is referred to as a \emph{reversible circuit} \cite{ref29, ref30, ref32, reversibleCompile}.

Because only the global unitary operator of a quantum circuit matters in our discussion, for simplicity we abstract away gate implementation details and denote
an $n$-qubit quantum circuit $C$ as a 4-tuple $C=(d,m,k,U)$ for $n=d+k$, where
\begin{itemize}
\item $d$ is the number of data qubits, specifically, $q_0, \ldots, q_{d-1}$,
\item $m$ is the number of measured qubits, specifically, $q_0, \ldots, q_{m-1}$,
\item $k$ is the number of non-data qubits, and
\item $U$ is the global unitary operator of $C$ of size $2^{n}\times 2^{n}$.
\end{itemize}
Note that above we make the data qubits and measured qubits follow the same order.
This arrangement is without loss of generality as the qubits can be reordered by swap gates.

Given a quantum circuit $C=(d,m,k,U)$ with the data qubits prepared at initial state $\psi$, we denote the probability of $C$ collapsing to state $|t\rangle$ upon a measurement on the measured qubits under the computational basis as $P(t|\psi,C)$, which can be calculated by
\begin{equation}
P(t|\psi,C) = \sum_{i=g\cdot t}^{g\cdot t+g-1} \left|\sum_{j=0}^{2^n-1} u_{i,2^k\cdot j}\cdot a_j\right|^2,
\label{eq3}\end{equation}
\noindent where $g=2^{(d+k-m)}$ and $u_{x,y}$ denotes the $(x,y)^\mathrm{th}$ entry of $U$.

We remark that a matrix can be represented numerically or algebraically, and explicitly or implicitly for different computation choices.
In this work, we adopt an algebraic and implicit approach to matrix representation and manipulation based on \cite{SliQSim,SliQEC}. 

\subsection{Algebraic and Implicit Representation of Matrices}

Following \cite{AlgebraicRepresentation}, we represent a complex number $\alpha \in \mathbb{C}$ algebraically by
\begin{equation}\label{eq4}
\alpha = \frac{1}{\sqrt{2}^k}(a\omega^3+b\omega^2+c\omega+d),
\end{equation}
\noindent where $\omega=e^{i\pi/4}$ and $a,b,c,d \in \mathbb{Z}$. 
In \cite{SliQEC}, the bit-slicing technique \cite{SliQSim} is applied to represent a $2^n \times 2^n$ complex matrix by one integer-type variable for the $k$-coefficient and $4r$ $2n$-variable BDDs for the $a$-, $b$-, $c$-, $d$-coefficients, assuming an integer is encoded with a bit-vector of size $r$. 
Let the BDDs of the $i^{\rm th}$ bit of the $a, b, c, d$-coefficient Boolean matrices be $F^{ai}, F^{bi}, F^{ci}, F^{di}$, respectively, for $i = 0, \ldots, r-1$. 
In the sequel, we do not distinguish among $F^{ai}, F^{bi}, F^{ci}, F^{di}$ and simply use $F$ to refer to every individual of these BDDs.
Essentially, such a function $F$ of $2n$ variables implicitly represents a $2^n \times 2^n$ Boolean matrix, whose rows and columns are indexed by the output qubits and input qubits, respectively, in our context.
For qubit $q_i$, two variables $x_i$ and $y_i$ are introduced to encode its corresponding output qubit and input qubit, respectively, for the Boolean matrix.
The variables $X =\{x_0, \ldots, x_{n-1}\}$ and $Y=\{y_0, \ldots, y_{n-1}\}$ are referred to as the row-variables and column-variables, respectively.
Effectively, $F$ is a function over variables $X \cup Y$.
That is, the $(i,j)^\mathrm{th}$ entry of the Boolean matrix corresponds to the Boolean value of $F$ for $X$ and $Y$ being substituted with the $n$-bit binary numbers of $i$ and $j$, respectively. 
In addition to the implicit matrix representation using Boolean functions, the multiplication of matrices can be done effectively through Boolean function manipulations as shown in \cite{SliQEC}.

\section{Partial Equivalence Problem}\label{sec:PECProblem}
In this section, we formally define partial equivalence between two quantum circuits, and compare it to other equivalence notions in the literature.

\subsection{Problem Statement}
The partial equivalence relation is formally defined as follows.

\begin{Definition}[Partial Equivalence]\label{Def4}
Two circuits $C_1=(d,m,k,U_1 )$  and $C_2=(d,m,k,U_2 )$ are \emph{partially equivalent}, denoted by $C_1\sim C_2$, if
\begin{equation} 
P(t|\psi,C_1) = P(t|\psi,C_2) 
\label{eq6}\end{equation} 
holds for any state vector $\psi$ of size $2^d$ and any $t=0,1,...,2^m-1$.
\end{Definition}
Note that, without loss of generality, we assume two circuits have the same number of ancilla qubits because we can always add dummy ancilla qubits to the one with fewer ancilla qubits. 
The partial equivalence checking problem can be stated as follows.
\begin{Problem}[Partial Equivalence Checking]\label{Problem1}
Given two circuits $C_1=(d,m,k,U_1 )$ and $C_2=(d,m,k,U_2)$, we are asked to check whether $C_1$ and $C_2$ are partially equivalent.
\end{Problem}

Partial equivalence can be applied to check the conformance between two quantum circuits that compute the same function but may produce different output states.

We remark that, in prior work \cite{ref25}, the notion of m-equivalence and q-equivalence are defined. 
For m-equivalent circuits, it requires that all qubits have fixed initial states, and it allows that we only measure on part of the qubits. 
Two circuits are m-equivalent if they have the same probability distribution of measurement outcomes. 
That is, m-equivalence is a special case of partial equivalence when we set $d=0$ for the circuits under verification.
On the other hand, for q-equivalence, ancilla qubits are allowed. 
Besides, the output qubits are divided into two groups, the main output qubits and the garbage qubits. 
It is required that if we measure the garbage qubits, the quantum states of the main output qubits cannot be affected by the measurement outcomes of the garbage qubits. 
Two circuits are q-equivalent if the output states of the main output qubits between the two circuits are the same for any input state.

\begin{figure}[t]
\centerline{\includegraphics[width=220pt]{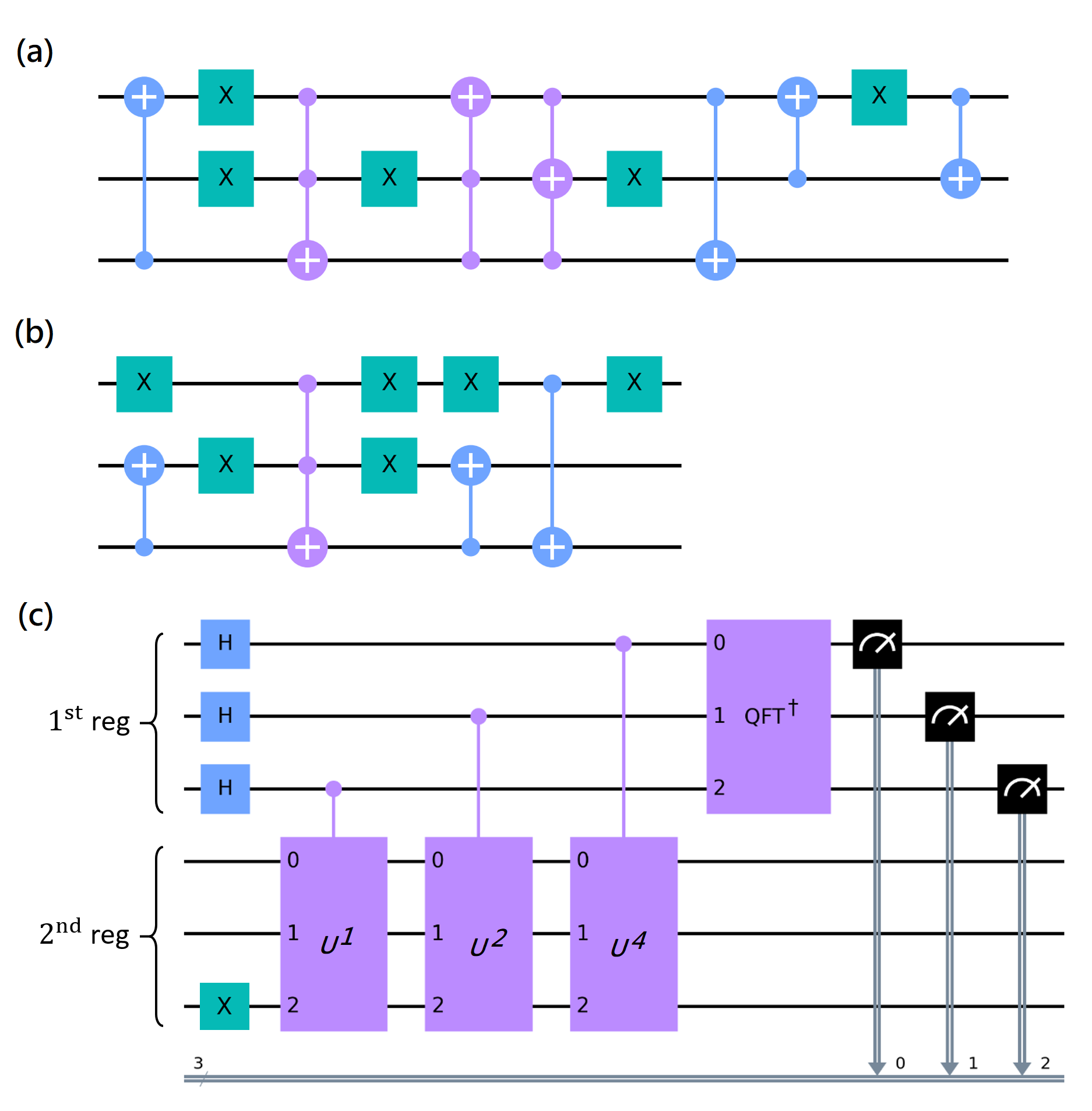}} 
\caption{\textbf{(a)} A reversible circuit of period 4 and with input state $|1\rangle$. 
\textbf{(b)} Another reversible circuit of period 4 and with input state $|1\rangle$. 
\textbf{(c)} The block diagram of the period-finding (PF) quantum algorithm. 
For two PF instances, one with $U$'s in (c) being substituted with the circuit in (a) and the other with $U$'s in (c) being substituted with that in (b), they are partially equivalent under $d=m=3$, i.e., the upper (resp. lower) 3 qubits in (c) are data qubits (resp. ancilla qubits fixed to initial state $|0\rangle$).} \label{PeriodFindingExample}
\end{figure}

To demonstrate the usefulness of partial equivalence, we take 
the period-finding quantum algorithm \cite{PF} as an example.
In Fig.~\ref{PeriodFindingExample} (a), the circuit computes the function
\begin{equation}
    U_1|x\rangle=
    \begin{cases}
        |3x\;\rm{mod}\;5\rangle, & \text{if $x<5$,} \\
        |x\rangle, & \text{otherwise.} 
    \end{cases}\qquad\qquad\;
\end{equation}
In Fig.~\ref{PeriodFindingExample} (b), the circuit computes the function 
\begin{equation}
    U_2|x\rangle=
    \begin{cases}
        |3(x+1)-1\;\rm{mod}\;5\rangle, & \text{if  $x<5$,} \\
        |x\rangle, & \text{otherwise.} 
    \end{cases}
\end{equation}
It can be seen that the $U_1$ and $U_2$ have the same period 4 (for $U_1$ mapping $|1\rangle \mapsto |3\rangle  \mapsto |4\rangle \mapsto |2\rangle \mapsto |1\rangle$ and $U_2$ mapping $|1\rangle \mapsto |0\rangle  \mapsto |2\rangle \mapsto |3\rangle \mapsto |1\rangle$).
For the circuit in Fig.~\ref{PeriodFindingExample} (c), let the first register as the data qubits and the second register as the ancilla qubits.
Let $C_1$ (resp. $C_2$) be the circuit of Fig.~\ref{PeriodFindingExample} (c) with the oracle blocks $U$ being substituted with function $U_1$ (resp. $U_2$).
Then $C_1$ and $C_2$ are partially equivalent under $d=m=3$ despite the fact that neither of total equivalence and q-equivalence holds and m-equivalence is not applicable  them.
The fact that $C_1$ and $C_2$ are not totally equivalent is immediate.
Also, we verify that the q-equivalence \cite{ref25} does not hold because the output state of the $1^{\rm st}$ register is affected by the $2^{\rm nd}$ register; the m-equivalence is not applicable because only the $2^{\rm nd}$ register is taken as the ancilla qubits. 
To the best of our knowledge, no previous tools can check this equivalence.

We observe that when the two circuits under verification have no ancilla qubits, the computation of their partial equivalence checking can be simplified.
It motivates the consideration of the following special case.
\begin{Problem}[Zero-Ancilla Partial Equivalence Checking]\label{Problem2}
Given two circuits $C_1=(d,m,0,U_1)$ and $C_2=(d,m,0,U_2)$, we are asked to check whether $C_1$ and $C_2$ are partially equivalent.
\end{Problem} 
For instance, consider the aforementioned $C_1$ and $C_2$ circuits. 
If we properly add some quantum gates to the beginning of the second register of $C_2$ (e.g., achieving the mapping $|x\rangle \mapsto |x-1\mod 5\rangle$ for $x<5$ and $|x\rangle \mapsto |x\rangle$  for $x\geq 5$), then $C_1$ and $C_2$ can remain partially equivalent even if all qubits are taken as data qubits.
In this case both circuits have no ancilla qubits (i.e., $k=0$).

\subsection{Comparison of Different Equivalences}
There have been several quantum circuit equivalences being studied. 
We compare them with the partial equivalence.

Besides the previously mentioned m-equivalence and q-equivalence, the most common definition of quantum circuit equivalence is \emph{total equivalence}, which requires that two circuits produce the same state vector (up to a global-phase difference) for any input state \cite{SliQEC,ref2,ref3,ref4,ref5,ref6,ref7,ref8,ref9,ref10,ref11,ref12}. 
That is, the output states of the two circuits should be numerically identical except for a scalar multiplication factor. 

Another well-known equivalence is functional equivalence of reversible circuits. 
In reversible circuits, they may also include ancilla qubits, and some qubits may be discarded in the end (i.e., we do not care about their measurement results). 
Some may even include don't-care conditions, which means that we do not care about all of or part of the measurement results under some input states \cite{ref29}.

Though partial equivalence looks similar to functional equivalence of reversible circuits and m-equivalence, we note that partial equivalence is much more complex, because measurement is a non-linear operation and does not have superposition property. 
That is, even if we know
\begin{equation}
P(t|\psi_i,C_1) = P(t|\psi_i,C_2)
\label{eq7}\end{equation}
for $i = 1, 2$,
we cannot infer 
\begin{equation}
P\left(t|(\alpha\psi_1+\beta\psi_2),C_1\right) = P\left(t|(\alpha\psi_1+\beta\psi_2),C_2\right).
\label{eq9}\end{equation}
Therefore, even with a brute-force search to enumerate the probability distribution under $\psi=e_{i,2^d}$ for all $i$, the partial equivalence relation is still not guaranteed. 
Some other approaches to checking partial equivalence are essential.



\begin{figure}[t]
\centerline{\includegraphics[width=220pt]{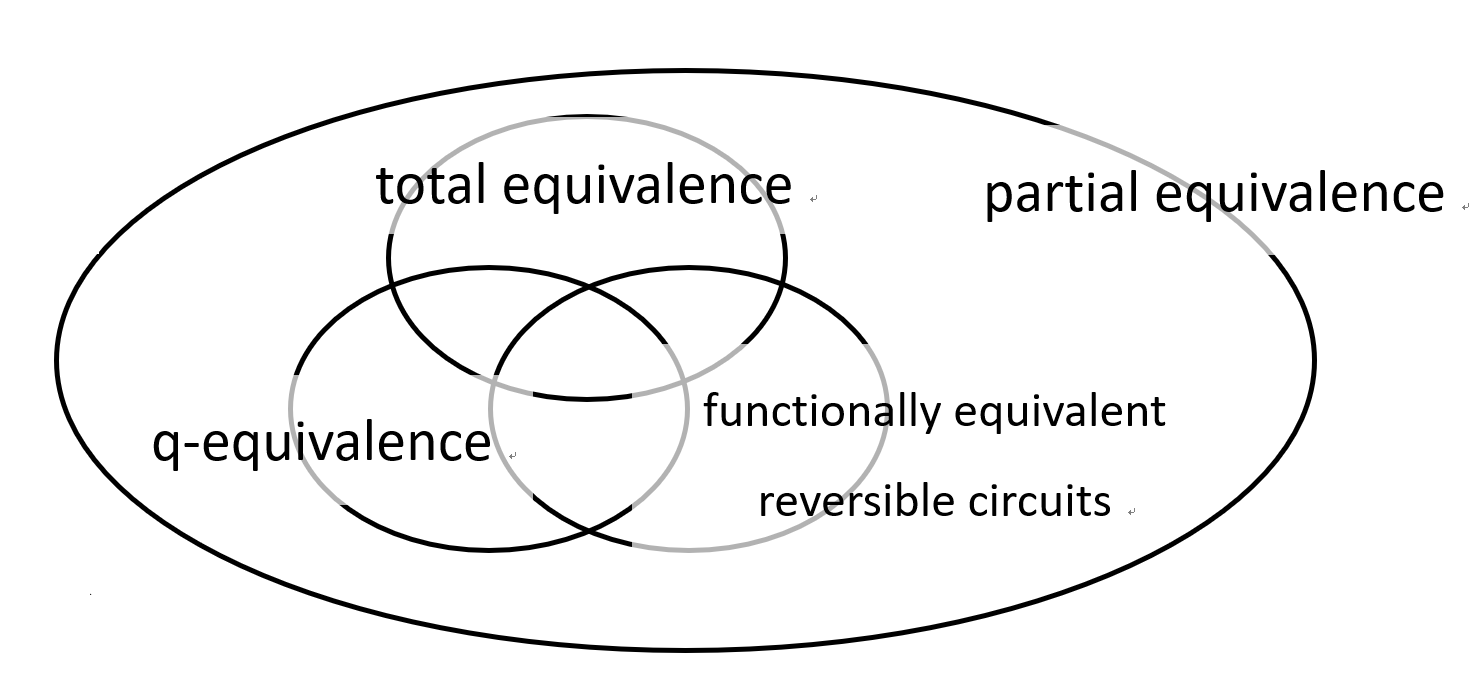}}
\caption{Relation between partial equivalence and other known equivalences. For m-equivalence, it is just the case that setting $d=0$ of partial equivalence and thus not displayed in the diagram.}\label{Fig3}
\end{figure}

We briefly summarize the main differences between the equivalence types mentioned above in Table~\ref{Comparing}. We further illustrate the relations between partial equivalence and other equivalences in Fig.~\ref{Fig3}. 
Note that partial equivalence subsumes total equivalence, functional equivalence of reversible circuits, and q-equivalence as these equivalences are special cases of partial equivalence. 
Specifically, it is evident that totally equivalent circuits must be partially equivalent. 
For reversible circuits, it can be derived from definition that $P(t|\psi, C)$ is either 0 or 1 for all $\psi=|0\rangle,|1\rangle,...,|2^d-1\rangle$, so Eq.~\eqref{eq9} holds. Therefore, functionally equivalent reversible circuits without don't care conditions must be partially equivalent. As for m-equivalence, it can be clearly seen that we can apply partial equivalence to m-equivalence by setting $d=0$. For q-equivalence, as the states of main output qubits between two circuits are identical, we must get the same probability distribution of the two circuits if we measure on the main output qubits. Thus, q-equivalent circuits must be partially equivalent. 

\renewcommand{\arraystretch}{1.3}
\begin{table*}[t]
\begin{center}
\caption{Comparison of Different Notions of Quantum Circuit Equivalence}
\begin{tabular}{|c||c|c|c|c|}
\hline
\textbf{equivalence type} & \textbf{equivalence form} & \textbf{inputs} & \textbf{outputs} & \textbf{special restrictions}\\
\hline
\hline

partial equivalence & measurement probability & allow ancilla & allow garbage & - 
\\ \hline
total equivalence & numerical quantum state & usually no ancilla & usually no garbage & - 
\\ \hline
\multicolumn{1}{|c||}{\tabincell{c}{functionally equivalent \\ reversible circuits}} & Boolean values & allow ancilla & allow garbage & restricted on reversible circuits
\\ \hline
m-equivalence & measurement probability & no ancilla & allow garbage & - 
\\ \hline
q-equivalence & numerical quantum state & allow ancilla & allow garbage & main output independent of garbage output
\\ \hline
\end{tabular}
\label{Comparing}
\end{center}
\end{table*}


\section{Necessary and Sufficient Conditions of Partial Equivalence}\label{sec:theorem}
In this section, we derive a necessary and sufficient condition for partial equivalence, which can be directly used to solve Problem \ref{Problem1} if the matrices of both circuits are explicitly given. For the special case with no ancilla qubits, we show that a simpler necessary and sufficient condition exists.


We define the following notation to locate some sub-matrices.
\begin{Definition}\label{Def3}
Given a circuit $C=(d,m,k,U)$, let $u_{p,q}$ represents the $(p,q)^\mathrm{th}$ entry of $U$ and $g=2^{(d+k-m)}$, then $v_{i,j}(U)$  of size $g$ is defined to be
\begin{equation} v_{i,j}(U)=[u_{gi,2^{k}j},u_{gi+1,2^{k}j},...,u_{gi+g-1,2^{k}j} ]^{\rm{T}},
\label{eq5}\end{equation}
where $i=0,1,...,2^{m-1}$, $j=0,1,...,2^{d-1}$. An example for $d=m=k=1$ is illustrated as follows. In addition, we use $e_{i,l}$  to represent the unit vector $[0,0,...,0,1,0,...,0]$  of length $l$ and the $i^{th}$  position ($i$ starts from 0) being 1.
\end{Definition}

\begin{figure}[H]
\centerline{\includegraphics[width=130pt]{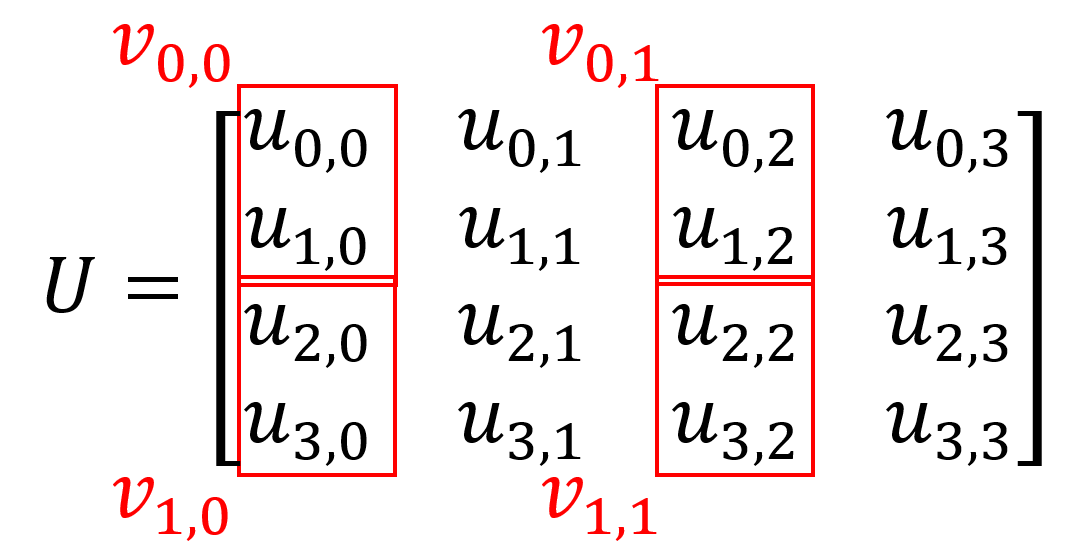}} 
\end{figure}

We remark that the definition of $\{v_{i,j}(U)\}$ is critical in deriving the necessary and sufficient condition for the partial equivalence.
We will show this point in the following section.

\subsection{Property of Partial Equivalence}

To solve Problem~\ref{Problem1}, we first derive the condition when $C_1 \sim C_2$ holds.

\begin{Theorem}[Partial Equivalence]\label{Thm1}
Given two circuits $C_1=(d,m,k,U_1)$  and $C_2=(d,m,k,U_2)$, then $C_1 \sim C_2$ if and only if
\begin{equation} v_{t,p}(U_1)^{\dagger}v_{t,q}(U_1)=v_{t,p}(U_2)^{\dagger}v_{t,q}(U_2) 
\end{equation}
holds for all $t=0,1,...,2^{m-1}$ and  $p,q=0,1,...,2^{d-1}$.
\end{Theorem}

\begin{proof}\label{Proof1}
Let the input state vector $\psi=[a_0,a_1,...,a_{(2^d-1)}  ]^{\rm{T}}$. The matrix $U_1$ and $U_2$ will respectively be applied on $\psi\otimes e_{0,2^{k}}$. Let's consider the output $|t\rangle=t_0 t_1...t_{m-1}$.
Let $g=2^{(d+k-m)}$  and $u_{1,x,y}$  denotes the $(x,y)$  entry of $U_1$, then

\begin{align}
\nonumber
&\qquad\quad P(t|\psi,C_1) = \sum_{i=gt}^{gt+g-1} \left|\sum_{j=0}^{2^d-1} u_{1,i,2^k\cdot j}\cdot a_j\right|^2  \\
\nonumber
&= \sum_{i=gt}^{gt+g-1} \left(  \sum_{p=0}^{2^d-1} \sum_{q=0}^{2^d-1} Re \left\{ (u_{1,i,2^k\cdot p}\cdot a_p)^*(u_{1,i,2^k\cdot q  }\cdot a_q)\right\}\right) \\
\nonumber
&= \sum_{p=0}^{2^d-1} \sum_{q=0}^{2^d-1} Re \left\{ a_p^* a_q \sum_{i=gt}^{gt+g-1}  u_{1,i,2^k\cdot p}^* \cdot u_{1,i,2^k\cdot q} \right\}\\
&= \sum_{p=0}^{2^d-1} \sum_{q=0}^{2^d-1} Re \left\{ a_p^* a_q \cdot v_{t,p}(U_1)^{\dagger}v_{t,q}(U_1) \right\},
\end{align}

\noindent where $Re\left\{x\right\}$  denotes the real part of the complex number $x$. Similarly, we get
\begin{equation}
P(t|\psi,C_2) = \sum_{p=0}^{2^d-1} \sum_{q=0}^{2^d-1} Re \left\{ a_p^* a_q \cdot v_{t,p}(U_2)^{\dagger}v_{t,q}(U_2) \right\}.
\end{equation}

If $C_1 \sim C_2$, then $P(t|\psi,C_1 )=P(t|\psi,C_2 )$  must hold for all $\psi$, so $v_{t,p}(U_1)^{\dagger}v_{t,q}(U_1)=v_{t,p}(U_2)^{\dagger}v_{t,q} (U_2)$  must hold for all $p$ and $q$. We also need the above equality hold for any $t=0,1,...,2^{m-1}$. Thus,
\begin{align}
C_1 \sim C_2 &\Leftrightarrow P(t|\psi, C_1 )=P(t|\psi, C_2 )   \;\forall t \forall \psi \nonumber\\
&\Leftrightarrow v_{t,p}(U_1)^{\dagger}v_{t,q}(U_1)=v_{t,p}(U_2)^{\dagger}v_{t,q}(U_2)  \;\forall t\forall p\forall q \nonumber.
\end{align}
\end{proof}

As mentioned before, brute-force search does not work for partial equivalence. Therefore, Theorem \ref{Thm1} provides the first way to formally verify the partial equivalence relation.

From Theorem \ref{Thm1}, we can see that whether $C_1 \sim C_2$ only depends on the $v_{i,j}(U)$’s of two matrices, and is not related to other entries. 

\subsection{Property of Zero-Ancilla Partial Equivalence}

For the case that both circuits have no ancilla qubits, we provide another necessary and sufficient condition for partial equivalence as follows.

\begin{Theorem}[Zero-Ancilla Partial Equivalence]\label{Thm2}
Given $C_1=(d,m,0,U_1)$ and $C_2=(d,m,0,U_2)$,
then $C_1 \sim C_2$ if and only if $U_1U_2^{-1}$ is in the following form.
\begin{equation}
    U_1U_2^{-1}=
    \begin{bmatrix}
      U_{0,0} &         &         &                  \\
              & U_{1,1} &         &                  \\
              &         &  \ddots &                  \\
              &         &         & U_{2^m-1,2^m-1} 
    \end{bmatrix}, \label{eq:miter}
\end{equation}
where each $U_{i,i}$ is a $2^{d-m} \times 2^{d-m}$ square matrix, and the other entries in $U_1U_2^{-1}$ not labeled are 0.
\end{Theorem}

\begin{proof}\label{prf2}
Firstly we show a transformed problem. By definition, $C_1 \sim C_2$ if and only if $P(t|\psi,C_1) = P(t|\psi,C_2)$ for any $\psi$ and $t$. Let $U_1\psi=\phi_1$, $U_2\psi=\phi_2$, and then we have $U_1U_2^{-1} \phi_2=\phi_1$. Because $U_1$ and $U_2$ are both reversible matrices, $\phi_1$ and $\phi_2$ can be any vector of size $2^d$ (and satisfying $\norm{\phi_1}= \norm{\phi_2}=1$). Thus, $C_1 \sim C_2$ if and only if $P(t|\phi_2,C_1C_2^{-1}) = P(t|\phi_2,I)$ for any $\phi_2$ and any $t$, where $C_1C_2^{-1}$ is the circuit with unitary matrix $U_1U_2^{-1}$, and $I$ is the identity circuit that directly sends the input to the output.

We will use the transformed problem in the following proof.\vspace{0.5em}

\noindent(i) the ``only if " part:

If $C_1 \sim C_2$, $P(t|\phi_2,C_1C_2^{-1}) = P(t|\phi_2,I)$ must hold for any $\phi_2$ and any $t$. We firstly consider $\phi_2=e_{0,2^d}$. We can easily find that
\begin{equation}
    P(t|e_{0,2^d},I)=
    \left\{
        \begin{aligned}
             & 1, \;\mathit{if} \; t=00...0 \\ 
             & 0, \;\mathit{else}
        \end{aligned}
    \right.    .
\end{equation}

Let $U=U_1 U_2^{-1}$ and $u_{x,y}$ denotes the $(x,y)$ entry of $U$. To make $P(00...0|e_{0,2^d},C_1C_2^{-1})=1$, $u_{i,0}$ should be 0 for all $i\geq 2^{d-m}$.

Following the same method, now we generally consider $\phi_2=e_{j,2^d}$, where $j=0,1,...,2^{d-1}$. Let $g=\lfloor j/2^{d-m}\rfloor$, then
\begin{displaymath}
\begin{aligned}
    &\;\;\;\;\;\;P(t|e_{j,2^d},I)=
    \left\{
        \begin{aligned}
             & 1, \;\mathit{if} \; t=g \\ 
             & 0, \;\mathit{else}
        \end{aligned}
    \right.\;\;\;\;\;\;\;\; \\
    &\Rightarrow P(t|e_{j,2^d},C_1C_2^{-1})=
    \left\{
        \begin{aligned}
             & 1, \;\mathit{if} \; t=g \\ 
             & 0, \;\mathit{else}
        \end{aligned}
    \right.\;\;\;\;\;\;\;\; \\
    &\Rightarrow u_{i,j}=0 \;\; \forall i \notin \left[ 2^{d-m}g,\,2^{d-m}(g+1) \right).
\end{aligned}
\end{displaymath}
\noindent That is, $U_1U_2^{-1}$ must be in form Eq.~\eqref{eq:miter}.\vspace{0.5em}

\noindent(ii) the ``if " part:

If $U=U_1U_2^{-1}$ is in form Eq.~\eqref{eq:miter}, we firstly note that:
\begin{equation}
    U^{\dagger}U=
    \begin{bmatrix}
      U_{0,0}^{\dagger}U_{0,0} & & &                  \\
      & U_{1,1}^{\dagger}U_{1,1} & &                  \\
      & & \quad\ddots\quad &                         \\
      & & & \quad\ddots\quad
    \end{bmatrix} .
\end{equation}
Because $U=U_1U_2^{-1}$ is a unitary matrix, $U_{i,i}^{\dagger}U_{i,i}$ must be identity for all $i$. Thus, each $U_{i,i}$ must be a unitary matrix.

Now for any $\phi_2$, let $\phi_2=\left[a_0,a_1,...,a_{2^d-1}\right]^{\rm T}=\left[A_0,A_1,...,A_{2^m-1}\right]^{\rm T}$, where each $A_i$ is of length $2^{d-m}$. That is, we equally divide $\phi_2$ into $2^m$ groups. Then
\begin{equation}
P(t|\phi_2,C_1C_2^{-1})=\norm{U_{t,t}A_t}.
\end{equation}
Because $U_{t,t}$ is unitary, it is true that
\begin{equation}
\norm{U_{t,t}A_t}=A_t^{\dagger}U_{t,t}^{\dagger}U_{t,t}A_t=\norm{A_t}.
\end{equation}
Because $\norm{A_t}=P(t|\phi_2,I)$, $P(t|\phi_2,C_1C_2^{-1})=P(t|\phi_2,I)$ holds for all $\phi_2$ and $t$. Therefore, $C_1 \sim C_2$. 
\end{proof}


\section{Algorithms for Partial Equivalence Checking}\label{sec:alg}
In this section, we present algorithms to solve Problems~\ref{Problem1} and \ref{Problem2} based on the theorems established in Section~\ref{sec:theorem}. 
We emphasize that these algorithms can be realized by using the bit-sliced algebraic representation of matrices \cite{SliQEC}.

\subsection{Partial Equivalence Checking Algorithm}\label{subSecI}

If $U_1$ and $U_2$ are given in their explicit numerical form, we can easily extract $v_{i,j}(U)$'s and enumerate all $(t,p,q)$ triples to check whether $C_1 \sim C_2$ by Theorem \ref{Thm1}, with $O(2^{3d+k})$ steps of complex number multiplications. However, as $U_1$ and $U_2$ may be in an implicit form, we give another approach by using matrix entry manipulation and matrix multiplication, as shown in Algorithm~\ref{alg1}.

\begin{algorithm} [h] 
\small 
\caption{\emph{PEC($C_1$,$C_2$)}} 
\begin{algorithmic} [1] \label{alg1}
\REQUIRE 
    {
    $C_1$: $C_1=(d,m,k,U_1)$, \\
    \quad\;\;\;$C_2$: $C_2=(d,m,k,U_2)$
    }
\ENSURE
    {
    Whether $C_1 \sim C_2$ or not
    }
\STATE $extra := {\rm max}\{m-k, 0\}$ 
\STATE $ k := k+extra$
\STATE	$U_1 := U_1 \otimes I_{2^{extra}}$
\STATE $U_1' := U_1$
\STATE Equally divide the columns of $U_1'$ into $2^d$ parts. For each part, keep the leftmost column unchanged, and set the remains to 0.
\STATE Equally divide the rows of $U_1'$ into $2^m$  parts. For the $i^{th}$ part ($i$ starts from 0), right shift the contents for $i$ columns.
\STATE $U_1'' := U_1^{\dagger}U_1'$
\STATE Equally divide the rows of $U_1''$ into $2^d$ parts. For each part, keep the top row and set the other entries to 0.
\STATE Repeat step 3$\sim$8 for $U_2$ and get $U_2''$.
\RETURN whether $U_1''=U_2''$ or not
\end{algorithmic}
\end{algorithm}

To better illustrate steps 5 and 6, we note that after step 6, each $v_{i,j}(U_1)$ for different $i$ and $j$ should be separated into different columns, and all entries not storing values of $v_{i,j}(U_1)$ are 0. An example of $U_1'$ with $d=m=k=1$ after this step is shown as follows.

\begin{figure}[h]
\centerline{\includegraphics[width=130pt]{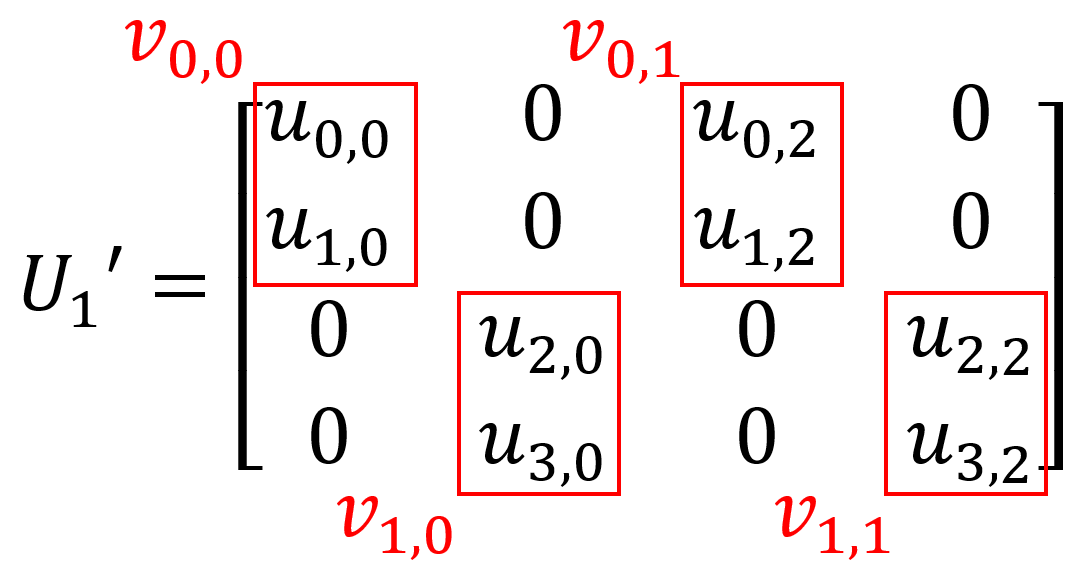}}
\end{figure}

To see the correctness of Algorithm~\ref{alg1}, we first consider the case $m \geq k$. It can be easily checked that after step 8, the $(2^kp,\:2^kq+t)^\mathrm{th}$ entry of $U_1''\:$will store the value of $v_{t,p}(U_1)^{\dagger}v_{t,q}(U_1)$, while the other entries are 0. Note that in step 2 we have added some ancilla qubits and now $k=m$. Therefore, we can easily check whether $U_1''=U_2''$ holds to determine whether $C_1 \sim C_2$.

For the case $m<k$, the $(2^kp,\:2^kq+t)^\mathrm{th}$ entry of $U_1''\:$still stores the value of $v_{t,p}(U_1)^{\dagger}v_{t,q}(U_1)$, but only for those with $t=0,1,...,2^{m-1}$. For $t=2^m,...,2^{k-1}$, the value of $v_{t,q}(U_1)^{\dagger}v_{t,q}(U_1)$ does not exist, but we can see that the $(2^kp,\:2^kq+t)^\mathrm{th}$ entry of $U_1''\:$will always be 0 because the entries in the ${(2^kq+t)}^{\rm{th}}$ column of $U_1'$ are all 0’s after step 6. That is, though we do not clear these entries in step 8, they are automatically set to 0. Therefore, we can still determine whether $C_1 \sim C_2$ by checking whether $U_1''=U_2''$.

We further point out that this can be done quite easily using bit-sliced algebraic representation of a matrix \cite{SliQEC}, as shown in Algorithm~\ref{alg2}. As introduced before, a matrix is represented by multiple BDDs, but here we use single $F_1$ and $F_2$ as representatives for all BDDs in $C_1$ and $C_2$ respectively for convenience. In steps 5, 7, 10, 13, we mean to do the same operation on each BDD, and in step 15, we need all corresponding BDDs of the two circuits to be the same.
\vspace{0.5em}

\begin{algorithm} [h]
\small
\caption{\emph{PEC\_BDD($C_1$,$C_2$)}}
\begin{algorithmic} [1] \label{alg2}
\REQUIRE 
    {
    $C_1$: $C_1=(d,m,k,U_1)$, \\
    \quad\;\;\;$C_2$: $C_2=(d,m,k,U_2)$
    }
\ENSURE
    {
    Whether $C_1 \sim C_2$ or not
    }
\STATE $extra := {\rm max}\{m-k, 0\}$ 
\STATE $ k := k+extra$
\STATE Add $extra$ row-variables and column-variables to $C_1$.
\FOR {$i=d,d+1,...,d+k-1$}
    \STATE $F_1 := F_1 |_{\overline{y_i}}$
\ENDFOR
\FOR { $i=0,1,...,m-1$ }
    \STATE $ F_1 := F_1 \land \overline{x_i \oplus y_{d+k-m+i} } $
\ENDFOR
\IF {$m<k$}
    \FOR {$i=d,d+1,...,d+k-m-1$}
        \STATE $F_1 := F_1 \,\land\, \overline{y_i}$
    \ENDFOR
\ENDIF
\STATE Apply the inverse circuit $C_1^{-1}$ on $F_1$ using the method in \cite{SliQEC}
\FOR {$i=d,d+1,...,d+k-1$}
    \STATE $F_1 := F_1 \land \overline{x_i} $
\ENDFOR
\STATE Repeat step 3 to 13 for $C_2$ and get $F_2$
\RETURN whether $F_1=F_2$ or not
\end{algorithmic}
\end{algorithm}

We use steps 4 to 10 in Algorithm~\ref{alg2} to replace steps 5 and 6 in Algorithm~\ref{alg1}. In steps 4 to 5, we divide the columns of $U_1$ into $2^d$ parts. For each part, we use cofactor operation to copy the leftmost columns to the other columns. In steps 6 to 9, we remove unwanted entries to ensure that each column only contains different $v_{i,j}(U_1)$'s. The remaining steps have similar functions as in Algorithm~\ref{alg1} but are rewritten in Boolean form. 

We can see that with the help of BDD, the operations can be simplified and are more efficient.
To be precise, two applications of inverse circuit operations $U^{-1}$ are required, and only $O(m+k)$ Boolean AND, OR, and cofactor operations are needed, which is linear to $m$ and $k$. As BDDs often give a good data compression, there may be a significant speedup for Algorithm~\ref{alg2}.

\subsection{Zero-Ancilla Partial Equivalence Checking Algorithm}

Similarly, we can directly use Theorem \ref{Thm2} to solve Problem \ref{Problem2} if $U_1$ and $U_2$ are explicitly given. Here we emphasize that it is also easy to check this property using bit-sliced algebraic representation of the matrix. The checking procedure is shown in Algorithm~\ref{alg3}. Again, we mean to do the same operation on each BDD in step 5, and we require all BDDs to be constant $false$ in step 6.

\begin{algorithm} [h]
\small
\caption{\emph{PEC\_BDD\_noAncilla($C_1$,$C_2$)}}
\begin{algorithmic} [1] \label{alg3}
\REQUIRE 
    {
    $C_1$: $C_1=(d,m,0,U_1)$, \\
    \quad\;\;\;$C_2$: $C_2=(d,m,0,U_2)$
    }
\ENSURE
    {
    Whether $C_1 \sim C_2$ or not
    }
\STATE Build the BDDs $F$ representing $C_1C_2^{-1}$ using the method in \cite{SliQEC}
\STATE $M := false$
\FOR {$i=0,1,...,m-1$}
    \STATE $ M := M \lor \left(x_i \oplus y_i\right) $
\ENDFOR
\STATE $ F := F \land M $
\RETURN whether $F = false$ or not
\end{algorithmic}
\end{algorithm}

We can see that in Algorithm~\ref{alg3}, we do not have to  build BDDs for $C_1$ and $C_2$ respectively, but rather merge them into one quantum circuit $C_1C_2^{-1}$.
As there may be some gates canceled out in $C_1C_2^{-1}$, Algorithm~\ref{alg3} may be more efficient than Algorithm~\ref{alg2}.

\section{Experimental Evaluation} \label{sec:experiment}
We implemented Algorithms~\ref{alg2} and \ref{alg3} in C/C++ under the framework of \textit{SliQEC} \cite{SliQEC}, and adopted CUDD \cite{ref28} package for BDD manipulation. 
All experiments were conducted on a server with Intel Xeon Silver 4210 CPU at 2.20 GHz and 128 GB RAM. 
The reported runtime refers to CPU time in seconds. 
The time-out (TO) limit is set to 600 seconds.

We conducted experiments of partial equivalence checking on benchmarks of period-finding instances, randomly generated instances, Grover search instances, and reversible circuit instances.
We also experimented total equivalence checking on randomly generated instances.
The results are discussed as follows.

\subsection{Checking Period-Finding Algorithm}\label{Exp_2}
We used the tool proposed in \cite{reversibleCompile} to implement the reversible circuits realizing $U|x\rangle=|ax \mod N\rangle$ for $x<N$ and $U|x\rangle=|a(x+1)-1 \mod N\rangle$ for $x<N$, which have the same period with input state $|1\rangle$. 
We then substituted the reversible circuits into the oracle used in the period finding algorithm. 
We let the first register consist of 3 qubits, and used one extra ancilla qubit to implement quantum Fourier transform. 
We then used Algorithm~\ref{alg2} to test the equivalence relation. 
As mentioned above, this type of equivalence cannot be verified by previous tools, and Algorithm~\ref{alg2} is the only way to check their equivalence relation. 
We have tested the $(a,N)$ pairs for all $a\in\{3,5,7,11\},N\in\{5,7,11,13\}$ and $a<N$. 
The parameters are set to $d=m=3$ and $k=\lceil \log_2N\rceil+1$.

The results are shown in Table~\ref{Exp_PeriodFinding}. 
All test cases yield correct answers. 
As these test cases are relatively small, we can see that the runtime is roughly proportional to the gate count. 
On the other hand, the memory usages are roughly the same because all test cases have similar numbers of qubits.

\begin{table}[t]
\begin{center}
\caption{Experimental Results on Period-Finding Benchmarks}
\begin{tabular}{|r|r|r|r|c|r|r|}
\hline
\bm{$a$} & \bm{$N$} & \tabincell{c}{\textbf{\#Gates} \\ \textbf{in} \bm{$C_1$}} & \tabincell{c}{\textbf{\#Gates} \\ \textbf{in} \bm{$C_2$}} & \textbf{time (s)} & \tabincell{c}{\textbf{memory} \\ \textbf{(MB)}}\\
\hline
3 & 5 & 115 & 87 & 0.180 & 12.874\\\hline
3 & 7 & 108 & 136 & 0.354 & 13.914\\\hline
3 & 11 & 227 & 325 & 0.794 & 14.496\\\hline
3 & 13 & 353 & 248 & 0.796 & 13.959\\\hline
5 & 7 & 143 & 164 & 0.441 & 13.922\\\hline
5 & 11 & 381 & 388 & 1.083 & 14.512\\\hline
5 & 13 & 353 & 507 & 0.806 & 13.173\\\hline
7 & 11 & 395 & 437 & 1.263 & 14.520\\\hline
7 & 13 & 479 & 241 & 1.136 & 14.787\\\hline
11 & 13 & 430 & 416 & 1.302 & 14.787\\\hline
\end{tabular}
\label{Exp_PeriodFinding}
\end{center}
\end{table}

\subsection{Checking Random Benchmarks}\label{Exp_3}
To further test the scalability of our algorithms, we generated some benchmarks with Clifford+T gates (H, S, T, CNOT gates) and 2-control Toffoli gates. 
The numbers of data qubits of the benchmarks range from $d=5,10,15,20,25,30,35,40,$ $50,60,70,80,90,100$. 
The number $m$ of measured qubits was fixed to $\lfloor0.5d\rfloor$, and the gate number was set to about $6.5d$. 
For each $d$, 20 groups of circuits were generated, and each group includes two conditions: ancilla qubits existing or non-existing. 
We used the latter condition to test Algorithm~\ref{alg3}, and both conditions to test Algorithm~\ref{alg2}.

\begin{figure}[b]
\centerline{\includegraphics[width=200pt]{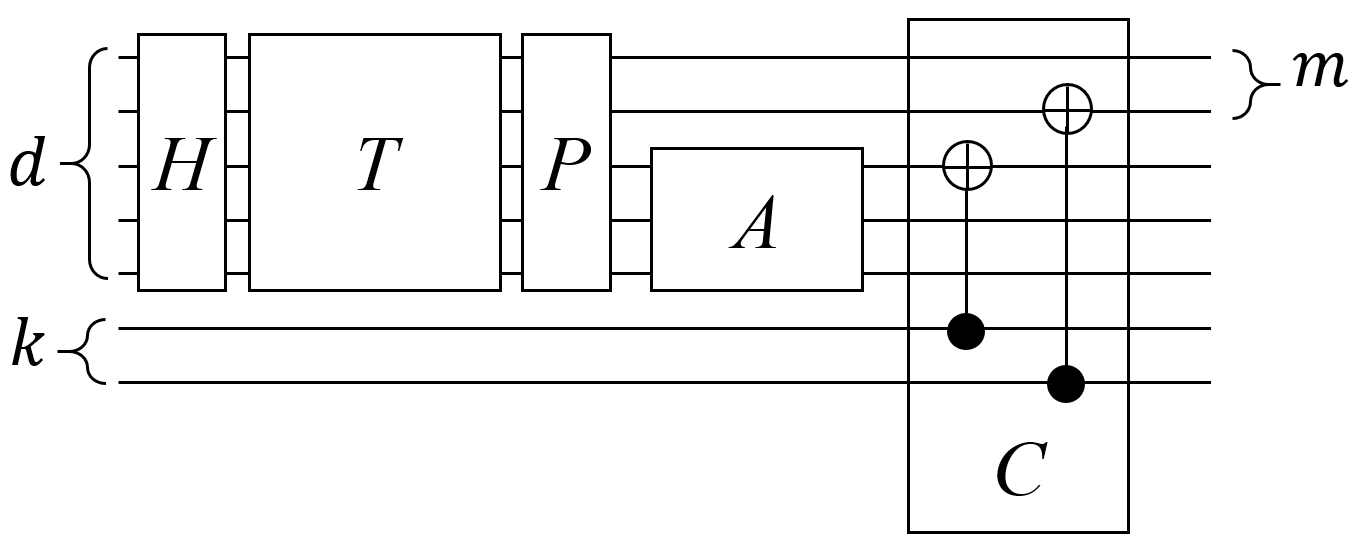}}
\caption{Structure of random benchmarks.}\label{RandomPartialStructure}
\end{figure}

The structure of the benchmarks is shown in Fig.~\ref{RandomPartialStructure}. 
Let the partially equivalent circuit pair be $C_1$ and $C_2$, both with $d$ data qubits and $m$ measured qubits. 
The circuits are divided into five parts: \textit{H} (H gates), \textit{T} (totally equivalent), \textit{P} (partially equivalent), \textit{A} (arbitrary) and \textit{C} (CNOT gates). 
If $C_1$ and $C_2$ are in the condition without ancilla qubits, only the first four parts are contained. 
Otherwise, they both contain all five parts.

In part \textit{H}, an H gate is firstly applied to each data qubit to impose superposition. 
In part \textit{T}, we generate a random sub-circuit with $d$ qubits and $3d$ gates to apply on both circuits, but all the Toffoli gates in $C_2$ are decomposed in the way proposed in \cite{ref33}. 
In part \textit{P}, we divide the data qubits into several groups, where each group may contain one or two adjacent qubits. 
For each group, we apply $X_1$ and $X_2$ on $C_1$ and $C_2$, respectively, where $X_1$ and $X_2$ are pre-generated subcircuits satisfying $X_1 \sim X_2$ when all qubits are set to data qubits and measured qubits.
$X_1$ and $X_2$ are exhaustively searched with up to five gates in total. 
In part \textit{A}, different random sub-circuits with $d-m$ qubits and $d-m$ gates are applied on $C_1$ and $C_2$. 
These gates trivially do not affect the measurement results of the measured qubits. 
In part \textit{C}, we use the ancilla qubits as the control bits to control the upper $d$ qubits. 
As ancilla qubits are initially set to 0, these CNOT gates do not affect the circuit.

Although the circuit generation is simple, there are no known circuit simplification methods exploiting partial equivalence. 
The generated circuits are used to test our algorithms. 
Furthermore, to validate the correctness of the benchmarks, we use the same generating method to generate some test cases with $3 \leq d \leq 9$. 
For each test case, we use \textit{Qiskit} \cite{ref20} to randomly set an initial state and get the final state vector numerically. 
We calculate for all possible outputs and make sure that $C_1$ and $C_2$ have the same probability distribution under this random initial state.

\begin{table*}[!t]
\begin{center}

\caption{Experimental Results on Random Partially Equivalent Circuits}
\begin{tabular}{|r|r|r|r|r|r|r|r|r|r|r|r|r|r|r|}
\hline
\multicolumn{1}{|c|}{\multirow{4}{*}{\bm{$d$}}} & \multirow{4}{*}{\bm{$m$}} & \multicolumn{8}{|c|}{\textbf{Without ancilla qubits}}&\multicolumn{5}{|c|}{\textbf{With ancilla qubits}} \\

\cline{3-15}
&&\multirow{3}{*}{\tabincell{c}{\textbf{\#Gates} \\  \textbf{in} \bm{$C_1$}}}&\multirow{3}{*}{\tabincell{c}{\textbf{\#Gates} \\ \textbf{in} \bm{$C_2$}}}&\multicolumn{3}{|c|}{\textbf{Algorithm 3}}&\multicolumn{3}{|c|}{\textbf{Algorithm 2}}&\multirow{3}{*}{\tabincell{c}{\textbf{\#Gates} \\  \textbf{in} \bm{$C_1$}}}&\multirow{3}{*}{\tabincell{c}{\textbf{\#Gates} \\ \textbf{in} \bm{$C_2$}}}&\multicolumn{3}{c|}{\textbf{Algorithm 2}}\\ 

\cline{5-10}\cline{13-15}

&&&&\textbf{time(s)}&\tabincell{c}{\textbf{memory} \\ \textbf{ (MB)}}&\textbf{\#TO}&\textbf{time(s)}&\tabincell{c}{\textbf{memory} \\ \textbf{ (MB)}}&\textbf{\#TO}&&&\textbf{time(s)}&\tabincell{c}{\textbf{memory} \\ \textbf{ (MB)}}&\textbf{\#TO}\\
\cline{1-15}
5 & 2 & 26.05 & 78.95 & 0.017 & 12.942 & 0 & 0.064 & 12.954 & 0 & 27.05 & 79.95 & 0.067 & 12.954 & 0\\\hline
10 & 5 & 51.05 & 157.00 & 0.050 & 12.942 & 0 & 1.880 & 25.077 & 0 & 52.65 & 158.60 & 1.951 & 24.766 & 0\\\hline
15 & 7 & 76.00 & 240.70 & 0.135 & 13.855 & 0 & 72.416 & 165.957 & 1 & 77.95 & 243.05 & 72.576 & 168.987 & 1\\\hline
20 & 10 & 102.65 & 308.75 & 0.364 & 14.180 & 0 & 117.528 & 170.217 & 14 & 105.85 & 312.00 & 129.006 & 167.451 & 13\\\hline
25 & 12 & 128.90 & 372.70 & 0.723 & 14.577 & 0 & - & - & - & 132.55 & 376.15 & - & - & -\\\hline
30 & 15 & 155.45 & 439.05 & 1.276 & 15.798 & 0 & - & - & - & 160.00 & 443.40 & - & - & -\\\hline
35 & 17 & 180.65 & 543.15 & 2.491 & 17.633 & 0 & - & - & - & 185.50 & 548.25 & - & - & -\\\hline
40 & 20 & 204.80 & 587.75 & 3.554 & 18.930 & 0 & - & - & - & 210.85 & 594.00 & - & - & -\\\hline
50 & 25 & 257.00 & 710.25 & 8.827 & 35.281 & 0 & - & - & - & 264.65 & 717.75 & - & - & -\\\hline
60 & 30 & 309.65 & 862.90 & 18.673 & 45.444 & 0 & - & - & - & 318.25 & 872.50 & - & - & -\\\hline
70 & 35 & 361.45 & 1037.45 & 55.950 & 78.525 & 0 & - & - & - & 372.10 & 1048.00 & - & - & -\\\hline
80 & 40 & 414.55 & 1157.00 & 105.611 & 122.198 & 3 & - & - & - & 426.00 & 1169.25 & - & - & -\\\hline
90 & 45 & 461.65 & 1328.75 & 229.473 & 144.761 & 6 & - & - & - & 475.80 & 1342.00 & - & - & -\\\hline
100 & 50 & 517.35 & 1464.15 & 297.401 & 160.311 & 11 & - & - & - & 532.70 & 1480.05 & - & - & -\\\hline
\end{tabular}
\label{RandomPartial}
\end{center}
\vspace{0.5em}
\begin{center}
\caption{Experimental Results on Functionally Equivalent Reversible Circuits} 
\begin{tabular}{|r|r|r|r|r|r|r|r|}
\hline
\multicolumn{8}{|c|}{ \textbf{Without ancilla qubits (Algorithm 3)} }\\
\hline
\multicolumn{1}{|c|}{\bm{$C_1$}} & \multicolumn{1}{|c|}{\bm{$C_2$}} & \multicolumn{1}{|c|}{\bm{$d$}} & \bm{$m$} & \textbf{\#Gates in} \bm{$C_1$} & \textbf{\#Gates in} \bm{$C_2$} & \textbf{time (s)} & \textbf{memory (MB)}\\
\hline
hwb4\_52 & hwb4\_49 & 4 & 4 & 11 & 17 & 0.0128 & 12.6730\\\hline
hwb6\_58 & hwb6\_56 & 6 & 6 & 42 & 126 & 0.0214 & 12.9434\\\hline
graycode6\_47 & graycode6\_48 & 6 & 6 & 5 & 5 & 0.0132  & 12.9434\\\hline
ham3\_103 & ham3\_102 & 3 & 3 & 4 & 5 & 0.0128 & 12.6730 \\\hline
urf1\_150 & urf1\_149 & 9 & 9 & 1517  & 11554  & 5.5048 & 15.4092\\\hline
urf2\_153  & urf2\_152 & 8 & 8 & 638 & 5030 & 1.1813 & 13.9059\\\hline
urf3\_156 & urf3\_155 & 10 & 10 & 2732 & 26468 & 25.6478 & 17.7562\\\hline
urf5\_159 & urf5\_158 & 9 & 9 & 499 & 10276 & 3.5319 & 15.4010\\\hline
urf6\_281  & urf6\_160 & 15 & 15 & 5102 & 10740 & 188.8090 & 112.1649\\\hline
alu-v1\_28 & alu-v1\_29 & 5 & 1 & 8 & 8 & 0.0101 & 12.9393\\\hline
alu-v2\_33 & alu-v2\_30 & 5 & 1 & 7 & 18 & 0.0130 & 13.0048\\\hline
alu-v3\_34 & alu-v3\_35 & 5 & 1 & 8 & 8 & 0.0110 & 12.9393\\\hline
alu-v4\_36 & alu-v4\_37 & 5 & 1 & 8 & 8 & 0.0127 & 12.9393\\\hline
\hline
\multicolumn{8}{|c|}{ \textbf{With ancilla qubits (Algorithm 2)} }\\
\hline
\multicolumn{1}{|c|}{\bm{$C_1$}} & \multicolumn{1}{|c|}{\bm{$C_2$}} & \multicolumn{1}{|c|}{\bm{$d$}} & \bm{$m$} & \textbf{\#Gates in} \bm{$C_1$} & \textbf{\#Gates in} \bm{$C_2$} & \textbf{time (s)} & \textbf{memory (MB)}\\
\hline
hwb5\_53  & hwb5\_300       & 5 & 5 & 55 & 101  & 3.8765 & 31.1869\\\hline
alu-v0\_26 & alu-bdd\_288 & 5 & 1 & 7 & 10  & 0.0146 & 12.9434\\\hline
ham7\_104 & ham7\_299 & 7 & 7 & 23 & 76  & 0.7321 & 16.2570\\\hline
ham15\_107 & ham15\_298 & 15 & 15 & 132 & 184  & 89.9754 & 203.5876\\\hline
nestedif2\_16\_396 & nestedif2\_16\_468 & 34 & 32 & 256 & 266  & 3.6910 & 37.0237\\\hline
\end{tabular}
\label{Reversible}
\end{center}
\end{table*}

The experimental results are shown in Table~\ref{RandomPartial}. 
The runtime and memory usage shown are averaged over the 20 test cases excluding those timed out. 
As seen, Algorithm~\ref{alg3} can easily scale up to tens of qubits. 
However, for $d>20$, Algorithm~\ref{alg2} can hardly finish the computation within the time limit. 
It may be because, for Algorithm~\ref{alg3}, $F$ keeps representing a unitary matrix of a quantum circuit almost during the whole process,
which may be more ``regular” for BDDs to represent. 
Moreover, due to the structure of $C_1C_2^{-1}$, some gates may be canceled out and the BDDs are more likely to be maintained as an identity matrix, as mentioned in \cite{SliQEC}. 
In contrast, the entries of the final matrix in Algorithm~\ref{alg2} contain the values of $v_{t,p}(U_1)^{\dagger}v_{t,q}(U_1)$, which can be arbitrary real numbers. 
Besides, we fix $m$ to $\lfloor0.5d\rfloor$ in our benchmarks, and the second step in Algorithm~\ref{alg2} may further increase the number of qubits. 
Thus, the runtime and memory usage of Algorithm~\ref{alg2} grow more rapidly than those of Algorithm~\ref{alg3}.

\subsection{Checking Reversible Circuits}\label{Exp_4}
As mentioned, functionally equivalent reversible circuits without don't-care conditions must be partially equivalent. 
For those with ancilla inputs, Algorithm~\ref{alg2} is needed. 
On the other hand, Algorithm~\ref{alg3} is sufficient if no ancilla inputs exist. 
We used some test cases from \textit{RevLib} \cite{ref31} and used Algorithms~\ref{alg2} and \ref{alg3} to test their partial equivalence to justify this point. 
The experimental results are shown in Table~\ref{Reversible}. 
It is justified that all functionally equivalent reversible circuits are indeed partially equivalent. 
The runtime and memory usage mainly grow proportional to the gate-count increase.

\subsection{Checking Grover's Algorithm}\label{Exp_5}
If reversible circuits are used as part of a quantum circuit, the entire quantum circuit may not be a reversible circuit anymore, and its verification becomes more complex. 
We took Grover's algorithm as an example for case study. 
In the experiment, we created a reversible circuit as the oracle function in Grover's algorithm. 
We also used the way proposed in \cite{ref32} to implement the same function with one ancilla qubit. Then, Algorithm~\ref{alg2} was applied to check the equivalence of the two circuit. 
The data qubit number $d$ and measured qubit number $m$ range from $d=m=6,10,14,18,22,26$. 

The experimental results are shown in Table~\ref{Grover}. 
The runtime and memory usage grow as $d$ increases.
It is reasonable because the circuit size is larger. 
We note that quantum circuits of Grover's algorithm with the same oracle function but implemented by different reversible circuits are q-equivalent.
In this experiment, we show that they are also partially equivalent.  
Apart from equivalent test cases, we also try to remove one gate from the second circuit to make them not partially equivalent. 
The runtime and memory usage for non-equivalent test cases are still close to equivalent test cases. 

\begin{table}[t]
\begin{center}
\caption{Experimental Results on Grover Search Benchmarks}
\begin{tabular}{|r|r|r|r|c|r|r|}
\hline
\bm{$d$} & \bm{$m$} & \tabincell{c}{\textbf{\#Gates} \\ \textbf{in} \bm{$C_1$}} & \tabincell{c}{\textbf{\#Gates} \\ \textbf{in} \bm{$C_2$}} & \textbf{time (s)} & \tabincell{c}{\textbf{memory} \\ \textbf{(MB)}}\\
\hline
6 & 6 & 35 & 37 & 0.257 & 13.201\\\hline
10 & 10 & 59 & 61 & 1.537 & 38.687\\\hline
14 & 14 & 83 & 85 & 9.613 & 153.371\\\hline
18 & 18 & 107 & 109 & 22.831 & 179.745\\\hline
22 & 22 & 131 & 133 & 49.149 & 192.492\\\hline
26 & 26 & 155 & 157 & 135.116 & 232.116\\\hline
\end{tabular}
\label{Grover}
\end{center}
\end{table}

\begin{table*}[!hb]
\begin{center}
\caption{Experimental Results on Random Totally Equivalent Circuits}
\begin{tabular}{|r|r|r|r|r|r|r|r|r|r|}
\hline
\multicolumn{1}{|c|}{\multirow{2}{*}{\bm{$d$}}} & \multicolumn{1}{|c|}{\multirow{2}{*}{\bm{$m$}}} & \multicolumn{1}{|c|}{\multirow{2}{*}{\textbf{\#Gates in} \bm{$C_1$}}} & \multicolumn{1}{|c|}{\multirow{2}{*}{\textbf{\#Gates in} \bm{ $C_2$}}} & \multicolumn{3}{|c|}{\textbf{Algorithm 3}} & \multicolumn{3}{|c|}{\textbf{SliQEC}} \\
\cline{5-10}
&&&& \textbf{time (s)} & \textbf{memory (MB)} & \textbf{\#TO}
& \textbf{time (s)} & \textbf{memory (MB)} & \textbf{\#TO}\\
\hline
10 & 10 & 40.00 & 118.40 & 0.038 & 12.955 & 0 & 0.028 & 13.000 & 0\\\hline
20 & 20 & 80.00 & 237.50 & 0.182 & 14.245 & 0 & 0.194 & 14.237 & 0\\\hline
30 & 30 & 120.00 & 377.60 & 0.685 & 15.699 & 0 & 0.640 & 15.285 & 0\\\hline
40 & 40 & 160.00 & 496.00 & 1.802 & 19.232 & 0 & 1.749 & 17.815 & 0\\\hline
50 & 50 & 200.00 & 649.40 & 3.612 & 28.114 & 0 & 4.722 & 28.233 & 0\\\hline
60 & 60 & 240.00 & 751.00 & 7.246 & 38.732 & 0 & 9.889 & 42.336 & 0\\\hline
70 & 70 & 280.00 & 887.60 & 12.165 & 45.731 & 0 & 10.086 & 38.610 & 0\\\hline
80 & 80 & 320.00 & 993.40 & 20.001 & 58.224 & 0 & 22.079 & 56.422 & 0\\\hline
90 & 90 & 360.00 & 1071.20 & 24.703 & 69.052 & 0 & 25.826 & 68.746 & 0\\\hline
100 & 100 & 400.00 & 1230.20 & 48.089 & 85.605 & 0 & 58.483 & 79.486 & 0\\\hline
\end{tabular}
\label{table4}
\end{center}
\end{table*}

\subsection{Checking Total Equivalent Circuits}\label{Exp_6}
Because totally equivalent circuits must be partially equivalent, under these circumstances we use totally equivalent circuits as the benchmarks to test Algorithm~\ref{alg3} and compare it against \textit{SliQEC} \cite{SliQEC}. 
Because Algorithm~\ref{alg3} is extended from \textit{SliQEC}, we expect that they would perform similarly.
We generated some benchmarks with qubit number $d=10,20,$ $30,40,50,60,70,80,90,100$, and set all qubits to be measured qubits for partial equivalence checking. 
The generating method is similar to how we generate partially equivalent pairs, except that only part \textit{H} and part \textit{T} are included to make them totally equivalent.
The results shown in Table~\ref{table4} confirm the expectation that Algorithm~\ref{alg3} and \textit{SliQEC} behave similarly in both runtime and memory usage on these instances.

\section{Conclusions}
In this work, we defined partial equivalence, which is concerned with observational equivalence with respect to partial measurement and a set of initial input states. 
We obtained the necessary and sufficient conditions of partial equivalence, and exploited them to develop partial equivalence checking algorithms.
Experimental results demonstrate the usefulness and strengths of our methods.
For future work, it may be interesting to explore partial equivalence in applications such as quantum circuit synthesis and quantum program compilation. 



\section*{Acknowledgments}
This work was supported in part by the Ministry of Science and Technology of Taiwan under grants MOST 110-2224-E-002-011 and MOST 111-2119-M-002-012, and MediaTek Research.
TFC was supported by the TSMC Scholarship.

\newpage

\bibliographystyle{IEEEtran}
\bibliography{reference}

\end{document}